\documentclass[10pt,apj,emulateapj]{emulateapj}

\newcommand{\fAGN}{$f_{\rm AGN}$}
\newcommand{\sig}{$\sigma$}
\newcommand{\mbh}{M$_{\rm BH}$}
\newcommand{\kms}{km s$^{-1}$}
\newcommand{\accrate}{$L_{\rm [OIII]}/M_{\rm BH}$}
\newcommand{\LOIII}{L$_{\rm [OIII]}$}
\newcommand{\sunit}{$L_{\odot}/M_{\odot}$}

\shorttitle{Environmental dependence of AGN activity. I}
\shortauthors{Choi et al.}

\begin{document}

\title{Environmental dependence of AGN activity. I.: the effects of host galaxy}

\author{Yun-Young Choi \altaffilmark{1}, Jong-Hak Woo\altaffilmark{2,3}, 
\& Changbom Park \altaffilmark{4}}
\altaffiltext{1}{Astrophysical Research Center for the Structure and
Evolution of the Cosmos, Sejong University,
Seoul 143-747, Korea, yychoi@kias.re.kr}
\altaffiltext{2}{Corresponding author, University of California Los Angeles, CA 90095-1547, 
woo@astro.ucla.edu}
\altaffiltext{3}{Hubble Fellow}
\altaffiltext{4}{Korea Institute of Advanced Studies, cbp@kias.re.kr}

\begin{abstract}
Using a large sample of local galaxies (144,940) with -17.5 $< M_{r} < $-22
and $0.025 < z < 0.107$, selected from Sloan Digital Sky Survey Data Release 5,
we compare AGN host galaxies with non-AGN galaxies at matched 
luminosity, velocity dispersion, color, color gradient, or 
concentration index, to investigate how AGN activity is related with galaxy properties.
The AGN sample is composed of Type II AGNs identified with flux ratios of
narrow-emission lines with signal-to-noise ratio $> 6$.
We find that the fraction of galaxies hosting an AGN(\fAGN) depends strongly
on morphology together with color, and very weakly on luminosity or
stellar velocity dispersion of host galaxies. In particular, \fAGN\ of early-type
galaxies is almost independent of luminosity nor velocity dispersion
when color is fixed. 
The host galaxy color preferred by AGNs is $u-r\approx2.0$ for early-type
hosts and $u-r=2.0 \sim 2.4$ for late-type hosts.
This trend suggests that AGNs are dominantly hosted by
intermediate-mass late-type galaxies because early-type galaxies with 
$u-r\approx2.0$ are very rare.
We also investigate how the accretion power varies with galaxy properties.
While the Eddington ratio ([OIII] line luminosity normalized by black hole mass) 
ranges over three orders of magnitude for both morphological types, 
late-type galaxies are the dominant hosts over all AGN power.
Among late-type galaxies, bluer color galaxies host higher power AGNs.
These results are consistent with a scenario that 
more massive and redder galaxies are harder to host AGNs since these galaxies 
already consumed gas at the center or do not have sufficient 
gas supply to feed the black hole. In contrast, intermediate-mass, 
intermediate-color, and more concentrated late-type galaxies are more likely to 
host AGNs, implying that perhaps some fraction of low-mass, blue, and less concentrated late-type galaxies
may not host massive black holes or may host very low-power AGNs.
\end{abstract}


\section{Introduction}

Since the discovery of active galactic nuclei (AGNs) at the centers of
nearby galaxies (Seyfert 1943), and at cosmological distance as highly
energetic optical and radio sources (Schmit 1963),
it has been a long-standing issue why some galaxies host AGN while a large fraction
of galaxies do not (e.g. Martini 2004; Jogee 2006).
Are there fundamental differences between non-AGN and AGN host
galaxies or are AGN host galaxies a random subset of normal galaxies?
Spatially resolved kinematics of very nearby galaxies revealed that
massive bulge-dominated galaxies host dormant supermassive black holes at
their center (Kormendy \& Gebhardt 2001; see Ferrarese \& Ford 2005 for a recent review),
implying the ubiquity of black holes in massive galaxies.
Hence, triggering AGN activity is related to the fuel supply mechanism,
by which dormant black holes become active.

In turn, as a feedback mechanism AGN activity can play an important role in
suppressing star formation and gas cooling in galaxy or galaxy cluster scales
as demonstrated in various semi-analytic models (Croton et al. 2006;
Robertson et al. 2006; Bower et al. 2006). Understanding how AGN activity is
related with galaxy properties and large scale environment may shed light
on the origin of the black hole -- galaxy connection, as observed
in the present-day universe (e.g. Ferrarese \& Merritt 2000; Gebhardt et al. 2000),
and its evolution (Woo et al. 2006, 2008; Peng et al. 2006; Younger et al. 2008).

Some galaxies with particular properties, indicative of an efficient fuel
supply mechanism, may present more AGN activity than other galaxies.
In a standard galaxy evolution picture, AGN activity is
naturally expected through galaxy merging and interactions, which can provide
gas to the center and feed the nuclear black holes
(e.g. Sanders et al. 1988; Hopkins et al. 2006).
Host galaxies of AGNs are expected to be different from non-AGN galaxies
in this scenario, and this picture may be relevant to explain highly energetic AGNs
at $z \sim $2--3 near the peak of quasar activity and star formation
(Hasinger et al. 2005).
Internal processes such as bar driven gas inflow (e.g. Combes 2003; see Kormendy \& Kennicutt
2004 for a recent review), turbulence in inter-stellar matter (e.g. Wada 2004), and stellar wind
(e.g. Ciotti \& Ostriker 2007) are also
considered as a gas supply mechanism, responsible for triggering AGN activity
even without invoking galaxy mergers.

However, it has not been clear whether AGN host galaxies have distinct properties
compared to non-AGN galaxies.
Various studies indicated that AGN host galaxies do not show excess of bar presence
or local density, compared to non-AGN galaxies (e.g. Combes 2003; cf. Maia et al. 2003).
HST imaging studies of quasar host galaxies at low redshift showed that
host galaxies have structures and stellar populations very similar to those of non-AGN galaxies (e.g. Bahcall et al. 1997; McLure et al. 2000; Dunlop et al. 2003).
Photometric and spectroscopic studies of radio-loud AGNs
showed that their host galaxies lie on the same fundamental
plane as quiescent galaxies (Bettoni et al. 2001; Barth et al. 2003; Woo et al. 2004, 2005).
In contrast, the presence of cold molecular gas and young stellar population in
quasar host galaxies suggested that these host galaxies are not typical early-type
galaxies (Scoville et al. 2003; Tadhunter et al. 2005). A recent study with deep
HST images revealed that some host galaxies have weak merging signatures,
indicating recent merging or interactions although majority of stars are relaxed
as in the case of quiescent galaxies (Bennert et al. 2008).

The Sloan Digital Sky Survey (SDSS) provides a large sample of non-AGN and AGN
host galaxies, enabling to overcome the difficulties caused by the relatively
small sample size in the previous studies (e.g. Ho et al. 1997; Hunt \& Malkan 2004).
Since the SDSS data became available, various effects on AGN activity have been
investigated, including host galaxy properties (Kauffmann et al. 2003; Rich et al. 2005;
Schawinski et al. 2007), local density, and large scale environment
(Kauffmann et al. 2004; Constantin \& Vogeley 2006; Constantin et al. 2008).
By studying a large sample of emission line galaxies, Kauffmann et al. (2003)
reported that early-type galaxies with AGN have bluer color than those without AGN,
indicating that AGN host galaxies are intermediate objects between the red sequence
and the blue cloud (see also, Schawinski et al. 2007). Similarly, it was reported that
early-type host galaxies of LINERS have younger stellar population than quiescent
early-type galaxies (Graves et al. 2007).

Since early-type and late-type galaxies are in different evolutionary stages
as demonstrated in the color-magnitude diagram  (e.g. Bell et al. 2004),
and the galaxy mass is the main driver of galaxy evolution as observed in galaxy
downsizing (Cowie et al. 1996; Treu et al. 2005; De Lucia et al 2006),
it is crucial to compare AGN and non-AGN galaxies with matched galaxy
morphology and mass scale.
Then various effects due to galaxy evolution can be cancelled out.
However, a detailed study with a large sample of matched host galaxy morphology and
mass has not been reported. For a magnitude limited sample of nearby $\sim$ 500
galaxies with detailed morphology information, Ho et al. (1997; see also Ho 2008
for a review) investigated the AGN fraction of each morphological type, reporting
$\sim$50\% AGN fraction in early-type (E--Sab) galaxies. The high AGN fraction
in this study is due to the fact that high spatial resolution ($<200$ pc)
was obtained and, hence, faint nuclear emission lines were detected without
strong contamination of host galaxies. With a much larger sample of emission line galaxies
from SDSS DR1, Kauffmann et al. (2003) reported that AGNs are preferentially found
in more massive galaxies and that host galaxies of high [OIII] luminosity AGNs have
younger stellar age than non-AGN galaxies. To achieve more detailed understanding
on triggering AGN activity, it is necessary to have morphological classification and
matched host galaxy properties in comparison, and proper normalization of AGN luminosity
by black hole mass. 

Using a large volume-limited sample with detailed galaxy properties,
we initiated an intensive project to investigate environmental effects
on AGN activity in three different scales: host galaxy, small scale environment,
and large scale environment. 
In the first paper of this series, we explore how AGN activity is related with
host galaxy properties using a large sample of SDSS galaxies.
Small and large scale environmental effects on AGN activity will be addressed
in the next paper (Y.-Y. Choi et al. 2009, in preparation).
We discuss our sample selection criteria in \S~2 and sample properties in \S~3.
In \S~4, we explore how AGN fraction changes depending on the host galaxy properties.
In \S~5, we present how AGN power is related with each host galaxy property.
Summary and discussion will be given in \S~6.
Throughout this paper, we adopt a flat $\Lambda$CDM cosmology with
$\Omega _{\rm m} = 0.27$,
and $\Omega_{\Lambda} = 0.73$. Magnitudes are given in the AB system.

\begin{deluxetable*}{lccccc}
\tablewidth{0pt}
\tablecaption{Volume-limited Samples}
\tablehead{
\colhead{Name}      &
\colhead{Magnitude} &
\colhead{Redshift}  &
\colhead{Distance}  &
\colhead{Number}    &
\colhead{$\bar d$}}
\startdata
sub6 & $-20.0>M_{\rm r}$       & $0.025<z<0.10713$& $74.6<R<314.0$& 80,478 & 5.56\\
sub5 & $-19.5>M_{\rm r}>-20.0$ & $0.025<z<0.08588$& $74.6<R<252.9$& 31,048 & 4.58\\
sub4 & $-19.0>M_{\rm r}>-19.5$ & $0.025<z<0.06869$& $74.6<R<203.0$& 16,772 & 4.18\\
sub3 & $-18.5>M_{\rm r}>-19.0$ & $0.025<z<0.05485$& $74.6<R<162.6$& 9,153  & 3.78\\
sub2 & $-18.0>M_{\rm r}>-18.5$ & $0.025<z<0.04374$& $74.6<R<129.9$& 4,932  & 3.41\\
sub1 & $-17.5>M_{\rm r}>-18.0$ & $0.025<z<0.03484$& $74.6<R<103.7$& 2,557  & 3.00\\
\enddata
\tablecomments{
Col. (1): name of subsamples.
(2): absolute magnitude range in the $r$-band.
(3): redshift range.
(4): distance range in units of $h^{-1} \rm Mpc$.
(5): the number of galaxies.
(6): mean separation of galaxies in units of $h^{-1} \rm Mpc$.
}
\end{deluxetable*}

\begin{figure}
\epsscale{1.2}
\plotone{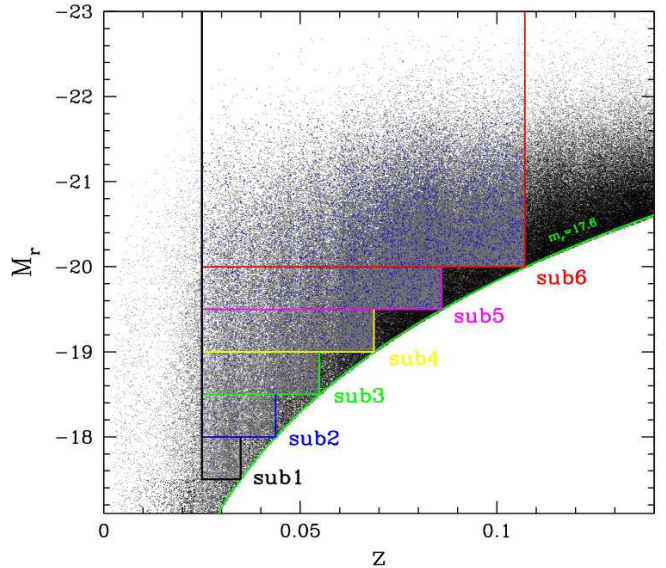}
\caption{The volume-limited sample.
The gray and blue dots in the whole area enclosed by rectangular boxes represent 
the total sample and the AGN sample, respectively.
Each subsample with various volume is defined with rectangular boxes.
The green curve, with Petrosian $r$-magnitude ($m_{\rm Pet}=17.6$) 
corresponds to the flux limit of the spectroscopic survey.}
\label{fig_Mz}
\end{figure}

\section{Data}

\subsection{Sample Selection}
We first selected galaxies with the $r$-band absolute magnitude 
$M_r < -20+5{\rm log} h$ (hereafter we drop the
$+5{\rm log} h$ term in the absolute magnitude) and $0.025<z<0.107$ from 
the large-scale structure sample, DR4plus,
using the New York University Value-Added Galaxy Catalogue 
(Blanton et al. 2005),
which is a galaxy subset (the Main galaxy sample) of the SDSS 
Data Release 5 (Adelman-McCarthy et al. 2007). 
Readers are referred to
Choi et al. (2007) for a detailed sample description.
The rest-frame absolute magnitudes of
individual galaxies are computed in the $r$-band, 
using Galactic reddening correction (Schlegel et al. 1998) and $K$-corrections
as described by Blanton et al. (2003). The mean evolution correction given by
Tegmark et al. (2004)
is also applied.
Our survey region, with angular selection function of more than 0.5,
covers 4464 deg$^2$ (see the survey boundaries in Figure 1 of Park et al. 2007).
To increase statistics toward low luminosity and low-mass scales,
we added five volume-limited samples selected from the same Catalogue, 
with fainter magnitude ranges down to $M_r = -17.5$.
The definitions of all 6 volume-limited subsamples are summarized in Table~1.

The total sample of 144,940 galaxies combined with all subsamples is presented with 
gray dots in rectangular boxes in Figure~\ref{fig_Mz}. The difference in the volumes of each 
subsample has been taken into account throughout our analysis.

\subsection {AGN Selection}

\begin{figure}
\epsscale{1.2}
\plotone{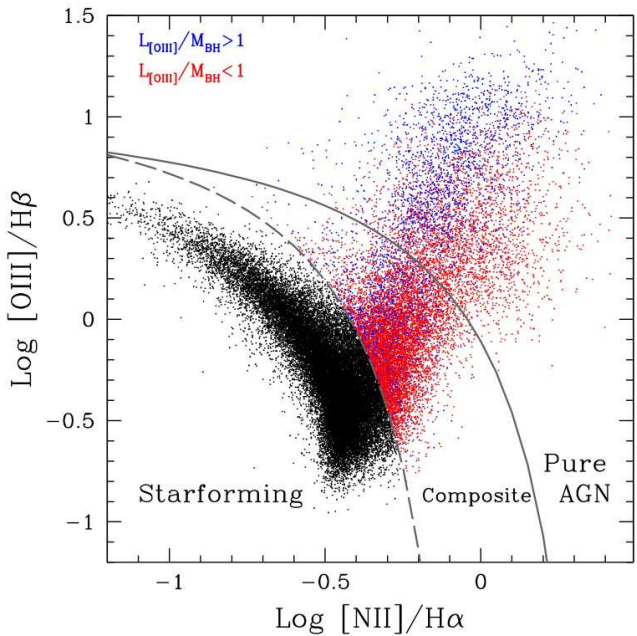}
\caption{Distribution of galaxies in the emission line flux ratios plane.
The AGN sample is defined by the solid curve (Kewley et al. 2006). 
To minimize contamination of star-forming galaxies, composite
objects below the solid line are excluded.
Blue dots represent high-power AGNs (\accrate$>1$ in $L_{\odot}/M{\odot}$)
while red dots represent low-power AGNs.
}
\label{fig_BPT}
\end{figure}

Since we will compare AGN and non-AGN galaxies with matched luminosity,
velocity dispersion, color, color gradient, and concentration, we selected
an AGN sample using Type II AGNs, for which host galaxy properties can be
determined in the same way as in non-AGN galaxies, without suffering further
complication due to the presence of a bright nuclear.
Type II AGNs can be separated from star forming galaxies 
based on the flux ratios of Balmer and ionization lines
(Baldwin, Phillips \& Terlevich, 1981; Veilleux \& Osterbrock 1987).
Figure~\ref{fig_BPT} presents the line ratios of emission-line galaxies, for which
all four emission lines ([OIII]$\lambda5007$, H$\beta$, [NII]$\lambda6583$, 
and H$\alpha$) are detected with signal-to-noise ratio (S/N) $\geq 6$.
Out of the 144,940 galaxies, 46,520 galaxies (32\%) satisfy the $S/N$ criterion.
Following Veilleux \& Osterbrock (1987) and other studies,
we used three classification schemes including low-ionization species. 
In practice, we selected AGNs using a conservative AGN definition from Kewley et al. (2006);
\begin{eqnarray}
0.61/({\rm log}([\rm NII]/H\alpha)-0.47)+1.19 &<&{\rm log}([\rm OIII]/H\beta),\nonumber\\
0.72/({\rm log}([\rm SII]/H\alpha)-0.32)+1.30&<&{\rm log}([\rm OIII]/H\beta),\nonumber\\
0.73/({\rm log}([\rm OI]/H\alpha)+0.59)+1.33&<&{\rm log}([\rm OIII]/H\beta)\nonumber.
\label{eq2}
\end{eqnarray}
We also used Kauffmann et al. (2003) demarcation line (dashed line in Fig.~\ref{fig_BPT})
to identify composite objects, which contain AGN as well as extended HII regions;
\begin{eqnarray}
0.61/({\rm log}([\rm NII]/H\alpha)-0.05)+1.30 &>&{\rm log}([\rm OIII]/H\beta),\nonumber\\
0.72/({\rm log}([\rm SII]/H\alpha)-0.32)+1.30&>&{\rm log}([\rm OIII]/H\beta),\nonumber\\
0.73/({\rm log}([\rm OI]/H\alpha)+0.59)+1.33&>&{\rm log}([\rm OIII]/H\beta). \nonumber
\end{eqnarray}

We excluded potential Type I AGNs. SDSS spectroscopic pipeline does not classify 
broad-line AGNs as `galaxies'. However, some narrow Type I AGNs can still be present
among galaxies. We assume galaxies with a H$\alpha$ emission line width larger than
$\sim$500 km s$^{-1}$ (FWHM) as Type I AGNs.
Only 412 objects were found in our sample and excluded.

Combining pure AGNs and composite objects, we defined an AGN sample, which
is composed of $11,521$ objects. The fraction of each class is shown 
in detail in Table~\ref{tab}. The AGN sample is 7.9\% of the total galaxy sample,
however we note that this fraction depends on the $S/N$ criterion on the emission lines.
For example, when we used $S/N>10$, the number of AGNs drops to 6,088.
Thus, the AGN fraction presented in this paper should be taken as a lower limit.

The emission line fluxes and stellar velocity dispersion were measured 
using an automated spectroscopic pipeline, {\tt specBS} version 5 
(D. Schlegel et al. 2009, in preparation).
The basic technique used in this pipeline is described by Glazebrook et al. (1998)
and Bromley et al. (1998). It determines radial velocity, redshift, and 
spectral classifications by fitting the spectra with SDSS-derived stellar templates 
as well as stellar templates drawn from the high-resolution spectra from the ELODIE
survey (Prugniel \& Soubiran 2001). The library of stellar templates is
constructed by applying the principle component analysis technique to a sample
of pure absorption-line galaxies.
Note that these emission line measurements are slightly different from 
those given by Tremonti et al. (2004) because of the difference between 
algorithms in subtracting stellar continuum before carrying out emission line fits.
However, only H$\beta$ line fluxes are somewhat affected by this 
stellar continuum correction, with 
$\sim 0.13\pm0.10$ dex difference
in log ([OIII]/H$\beta$) compared to Tremonti et al.
(the difference in log([NII]/H$\alpha$) is only  $0.02\pm0.02$ dex).
Thus, a small fraction of composite objects in our sample was classified as star-forming galaxies
in Kauffmann et al. (2003).
However, our AGN sample with $S/N>6$ is still more 
conservatively defined than that of Kauffmann et al.(2003) with $S/N>3$.

We adopt the [OIII] line luminosity ($L_{\rm[OIII]}$) as an accretion rate indicator,
after correcting for dust extinction by using the relation (Bassani et al. 1999);
\begin{equation}
L_{\rm [OIII]}= 
L_{\rm [OIII],obs} [({\rm H}\alpha/{\rm H}\beta)_{\rm obs}/({\rm H}\alpha/{\rm H}\beta)_{\rm o}]^{2.94},
\end{equation}
where an intrinsic Balmer decrement (H$\alpha$/H$\beta)_{\rm o} =3.1$ is adopted (Kewley et al. 2006).
Black hole mass (\mbh) was estimated from stellar velocity dispersion
assuming the \mbh-$\sigma$ relation (Tremaine et al. 2002);
\begin{equation}
{\rm log} M_{\rm BH}/M_{\odot} = 8.13 + 4.02 \times 
{\rm log} (\sigma/200~{\rm km~s}^{-1}),
\end{equation}
where a simple aperture correction to the stellar velocity
dispersion has been applied (Bernardi et al. 2003).
Then, the Eddington ratio is quantified by 
$L_{\rm [OIII]}$ normalized by \mbh, as an indicator of AGN power.

\section{Sample Properties}

To investigate how AGN activity is related with each host galaxy property,
it is crucial to match other galaxy properties in comparing AGN and non-AGN 
galaxies.
Morphology and mass are probably the two most fundamental parameters 
of galaxies, presenting the current evolutionary status.
By fixing morphology and mass,
we can avoid confusions resulting from the correlations of other
physical parameters with mass and morphology.
In this section, we discuss our morphology classification scheme (\S~\ref{sec_mor}),
the luminosity and velocity dispersion distributions (\S~\ref{sec_lsigma}), and
the Eddington ratio distributions of the AGN sample (\S~\ref{sec_OIIIMBH}).

\subsection{Morphology Classification}
\label{sec_mor}

\begin{figure}
\epsscale{1.2}
\plotone{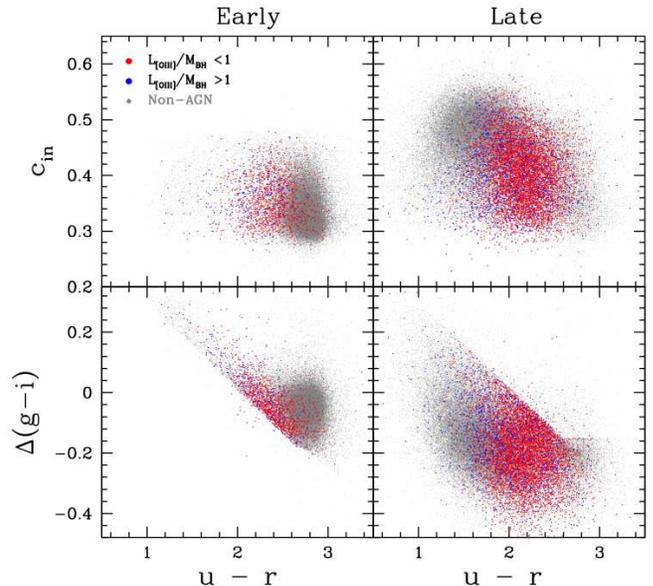}
\caption{Segregation between early-type ({\it left}) and 
late-type galaxies ({\it right}) in the $u-r$ color versus 
inverse-concentration index ({\it upper}) and $u-r$ color 
versus and $g-i$ color gradient ({\it lower}). Gray points are
non-AGN galaxies, and blue(red) points are host galaxies of high(low)-power AGNs.
[{\it See the electronic edition of the Journal for a color version of this figure.}]
}
\label{fig_morph}
\end{figure} 

Morphology classification is not straight forward especially for faint and
seeing-limited distant galaxies. To apply to a large sample of SDSS galaxies,
we adopted from Park \& Choi (2005) a classification tool in the color ($u-r$) 
versus color gradient ($\Delta (g-i)$) plane, where early-type and late-type 
galaxies are well separated.
The $g-i$ color gradient is defined as the color difference between 
the annulus with $0.5 R_{\rm Pet} < R <R_{\rm Pet}$ and 
the region with $R<0.5 R_{\rm Pet}$, 
where $R_{\rm Pet}$ is the $i$-band Petrosian radius.
The concentration index ($c_{\rm in}$) was also used as an auxiliary parameter,
which is defined as $R_{50}/R_{90}$, where $R_{50}$ and $R_{90}$
are the radii from the center of a galaxy containing 50\% and 90\% of 
the Petrosian flux in the $i$-band, respectively.
The boundaries between the two types in the three-dimensional parameter space
is determined in such a way that the classification best reproduces
the visual morphology classification: ellipticals/lenticular (early-type) and
spirals/irregulars (late-type). 

Using this classification tool, we divide our sample into two morphological groups.
Note that our morphology classification is
mainly based on color and color gradient, however, the reliability of this scheme
is $\sim$90\% based on the test with training samples, for which color-based
morphology was compared with morphology determined by eye 
(see Park \& Choi 2005 for details). 
For practical reasons, we will call these two groups as early-type and late-type galaxies.

Figure~\ref{fig_morph} shows our sample galaxies
in the $u-r$ color versus $c_{\rm in}$ plane (upper panels) 
and in the $u-r$ versus $\Delta(g-i)$ plane (lower panels).
Red (blue) dots represent low (high) power AGNs as in Figure~\ref{fig_BPT}
while gray dots denote non-AGN galaxies.
Note that the sharp boundaries shown in the $u-r$ versus $\Delta(g-i)$ 
plane are due to the morphology classification criteria and 
that the scatter across the classification boundary   
is caused by the concentration index constraint (Park \& Choi 2005).

Our morphology classification yields a significant fractions of early types 
with bluer color and positive color gradient (bluer inner part).
Among early-type galaxies, 54\% of AGN host galaxies is bluer than $u-r=2.4$, while
only 6\% of non-AGN galaxies have $u-r < 2.4$.
In the case of late-type galaxies, 17\% of AGN hosts are redder 
than $u-r=2.4$, similar to that (15\%) of non-AGN galaxies.
These results imply that galaxies with intermediate color have higher fraction
of AGN (see \S~\ref{subsec_color} for more discussion).

Table~\ref{tab} presents AGN fractions for each morphology group.
In the total sample, 7.9\% of galaxies host AGN, of which 83\% are late-type 
galaxies. 27\% of AGNs has high accretion rate, \accrate$>$1 ($L_{\odot}/M_{\odot}$)
and 84\% of these high-power AGNs are hosted by late-type galaxies. 
When we further exclude composite objects, 1.8\% of galaxies host AGN, 
of which 72\% are late-type galaxies.

\begin{deluxetable}{lcrrrr}
\tablewidth{0pt}
\tablecaption{Sample Statistics}
\tablehead{
\colhead{\% (Number)} & \colhead{All} &\colhead{Early}&
\colhead{Late}}    
\tablecolumns{4}
\startdata
All galaxies           &100.0 & 41.3\% (59,799) &58.7\% (85,141) \\ 
\\
pure AGN                & 1.8  & 27.5\% (718)    & 72.4\% (1,887) \\
composite               & 6.1  & 13.3\% (1,191)  & 86.7\% (7,722) \\
total AGN               & 7.9  & 16.6\% (1,909)  & 83.4\% (9,609) \\
\\
Non-AGN                 &73.5& 43.2 (45,981) & 56.8 (60,516)\\
\enddata
\label{tab}
\tablecomments{
Col. (1): class.
(2): fraction in total.
(3,4): early-type galaxy fraction and number in each AGN class.
(5,6): late-type galaxy fraction and number in each AGN class.
AGNs are classified based on the ratios of emission lines detected with $S/N>6$.
Note that non-AGN includes all galaxies excluding pure AGNs and composite objects
defined by emission line ratio without $S/N$ cut.}

\end{deluxetable}

\subsection{Luminosity and velocity dispersion}
\label{sec_lsigma}  

\begin{figure}
\epsscale{1.2}
\plotone{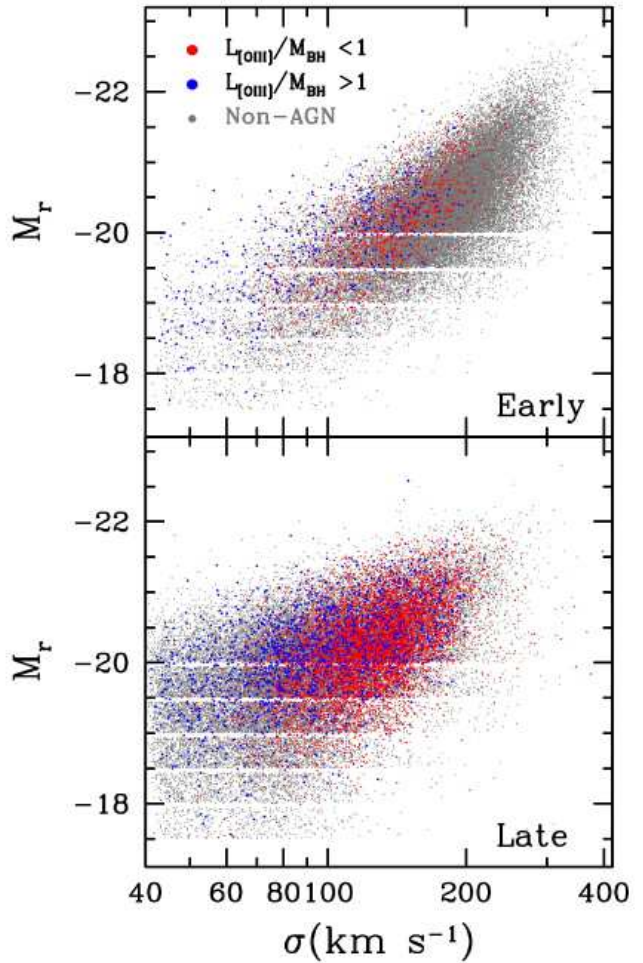}
\caption{Luminosity versus stellar velocity dispersion of early-type ({\it upper})
and late-type galaxies ({\it lower}). Gray dots represent non-AGN galaxies
while blue ({\it red}) dots represent high ({\it low})-power AGNs.
Volume-limited subsamples are displayed with a magnitude gap for clarification.
[{\it See the electronic edition of the Journal for a color version of this figure.}]}
\label{fig_msigma}
\end{figure}

Figure~\ref{fig_msigma} presents the relation between luminosity and 
stellar velocity dispersion (\sig) for non-AGN and AGN host galaxies.
For the stellar velocity dispersion, only spectra with mean $S/N$ per spectral pixel
greater than 10 were used. A simple aperture correction due to the finite size of
optical fiber has been applied to the velocity dispersion (Bernardi et al. 2003). 

It is already noticeable that the distribution of AGN host galaxies is different
from that of non-AGN galaxies in the sense that for given velocity dispersion, 
AGN host galaxies have higher luminosity than non-AGN galaxies,
in particular for early-type galaxies.
We will discuss in detail 
the relation between AGN activity and luminosity or velocity dispersion 
in \S~\ref{subsec_LV}. 
We note that at the same luminosity, 
late-type galaxies can have a systematically wide distribution in mass.
This may be due to the various levels of young stellar population fraction.
Thus, direct comparison between two morphological types is not trivial.

\subsection{Eddington ratio distribution}
\label{sec_OIIIMBH}

\begin{figure}
\epsscale{1.2}
\plotone{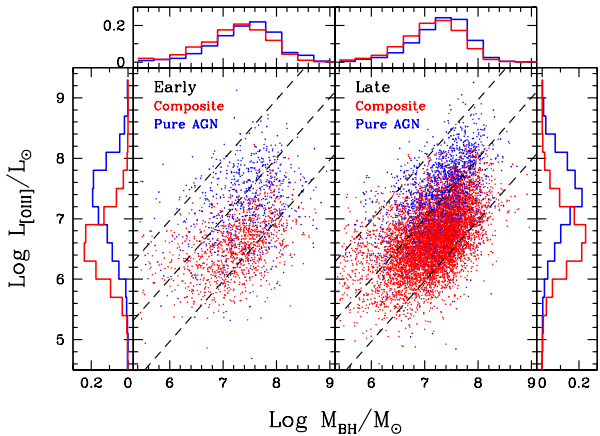} 
\caption{[OIII] luminosity versus \mbh\ for AGNs hosted by early-type
({\it left}) and late-type galaxies ({\it right}). 
Dashed lines from upper left to lower right indicate
$\sim$100\%, $\sim$10\%, and $\sim$1\% of the Eddington limit, respectively, 
assuming bolometric correction 3500 for $L_{\rm[OIII]}$ (with an uncertainty of $\sim$0.4 dex;
Heckman et al. 2004). 
We define high-power AGNs with \accrate $> 1$ ($L_{\odot}/M_{\odot}$),
corresponding to $>$10\% of the Eddington limit. Upper panels show relative 
fraction of AGNs as a function of \mbh.
[{\it See the electronic edition of the Journal for a color version of this figure.}]
}
\label{fig_OIIIMBH}
\end{figure}

Figure~\ref{fig_OIIIMBH} presents distributions of AGNs in 
the $L_{\rm[OIII]}$ versus \mbh\ plane.
Since \mbh\ correlates with the stellar velocity dispersion of
the spheroidal component, the derived \mbh\ of late-type galaxies suffers systematic
uncertainties. In practice, however, we cannot separate the bulge stellar
velocity dispersion for our sample of SDSS galaxies, and take the total velocity 
dispersion to estimate \mbh. Low \mbh\ values should be taken more cautiously.
First, \mbh\ estimates lower than $10^{6.3} M_{\odot}$ are not reliable since
velocity dispersion measurements lower than the instrumental resolution 
($\sim70$ \kms) are not robust. 
Second, there is a deficiency of objects with $L_{\rm [OIII]} \lesssim 10^{6}$$L_{\odot}$, 
presumably caused by the detection limit of the [OIII] line 
since weak [OIII] lines can be easily diluted by the host galaxy light.
Therefore, we will focus on AGNs with \mbh\ $>$ 10$^{6.7}$ $M_{\odot}$ (i.e. \sig\ $\sim 90$ \kms)
in the following analysis when host galaxy properties are related with AGN power.   

We take further caution due to the fact that disk galaxies can have 
shock-oriented extended emission line regions, which could mimic AGNs 
properties (Dopita \& Sutherland 1995; see discussion by Ho 2008).
The potential host galaxy contamination is unavoidable in the SDSS spectroscopy
with a 3$\arcsec$ diameter aperture, corresponding to several kpc 
for typical galaxies in the sample.
However, if the [OIII]/H$\beta$ ratio of extended
emission line region is different from the nuclear region, 
the contamination may not be severe (see Fehmers et al. 1994).

We split our AGNs into high and low-power subsamples at  
\accrate $= 1$ (\sunit), 
which corresponds to $\sim5-10$\% of the Eddington limit (top dashed line)
depending on the bolometric correction (Heckman et al. 2004). 
Figure~\ref{fig_OIIIMBH} shows that $L_{\rm [OIII]}$ 
varies roughly from the Eddington limit down to $\sim 10^{-2}$ of the Eddington luminosity (bottom dashed line).
However, at the lower \mbh ($<10^{6.7} M_{\odot}$), the Eddington ratio range becomes
narrower due to the incompleteness. 
Thus, we will use the high-power subsample (\accrate $> 1$) 
to test any $L_{\rm [OIII]}$-related selection effect
in the following analysis.

\section{Effects on the AGN fraction}
\label{sec_prop}

In this section, we examine how the fraction of galaxies hosting an
AGN is related with 
host galaxy properties. We define the AGN fraction (\fAGN) as 
the ratio between the number of AGN hosts and that of all galaxies
at fixed galaxy properties. We will investigate how \fAGN\ 
varies as a function of different galaxy properties:
luminosity and velocity dispersion (\S~\ref{subsec_LV}),
color (\S~\ref{subsec_color}), color gradient (\S~\ref{subsec_colorgrad}), and 
light concentration (\S~\ref{subsec_cindex}). 

\subsection{Luminosity and Velocity dispersion}
\label{subsec_LV}

\begin{figure}
\epsscale{1.2}
\plotone{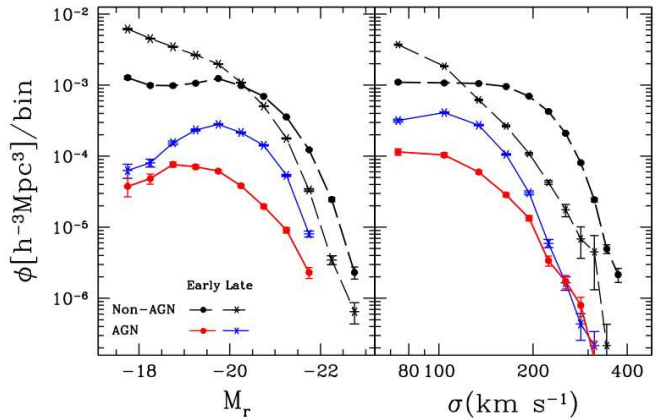}
\caption{The type-specific luminosity ({\it panel a}) and velocity dispersion
({\it panel b}) functions of non-AGN ({\it dashed lines}) and 
AGN host galaxies ({\it solid lines}), where the bin size of each plot 
is 0.5 mag and 30 km s$^{-1}$ 
for panels (a) and (b), respectively.
{[\it See the electronic edition of the Journal for a color version of this figure.]}
}
\label{fig_lf}
\end{figure}

\begin{figure*}
\epsscale{1.1}
\plotone{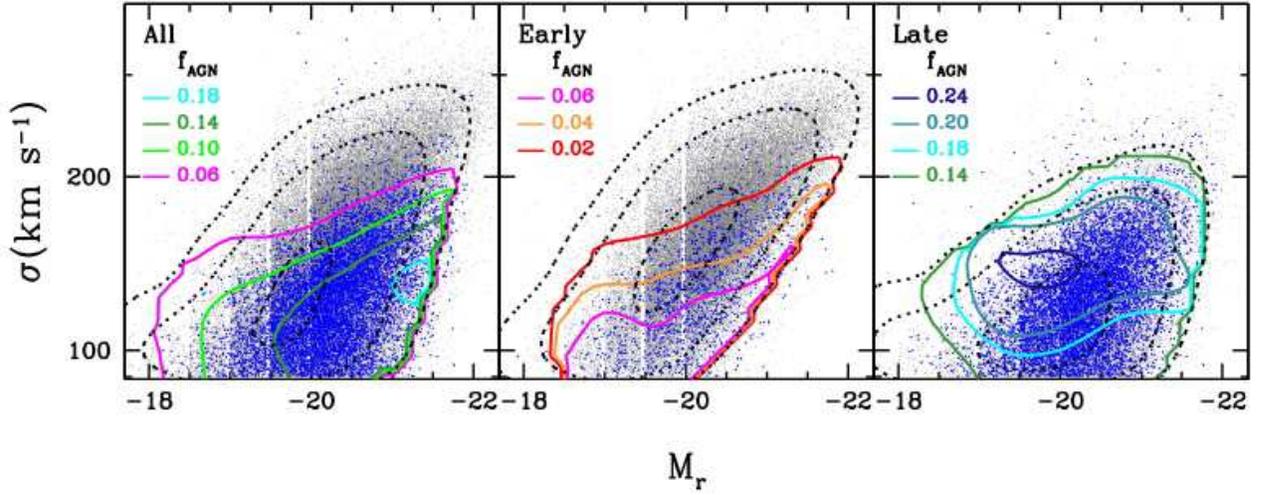}          
\caption{Distributions of AGN host ({\it blue dots}) 
and all galaxies in the sample ({\it gray dots}) 
in the luminosity versus velocity dispersion plane. The contours 
denote constant AGN fractions ({\it colored solid lines}) and galaxy number 
densities ({\it dotted lines}) where the bin sizes are 
$\Delta(log \sigma)=0.02$ and $\Delta({M_r})$=0.14.
The galaxy density contour levels are on logarithmic scales.
}
\label{fig_mcon}
\end{figure*}

\begin{figure*}
\epsscale{1.1}
\plotone{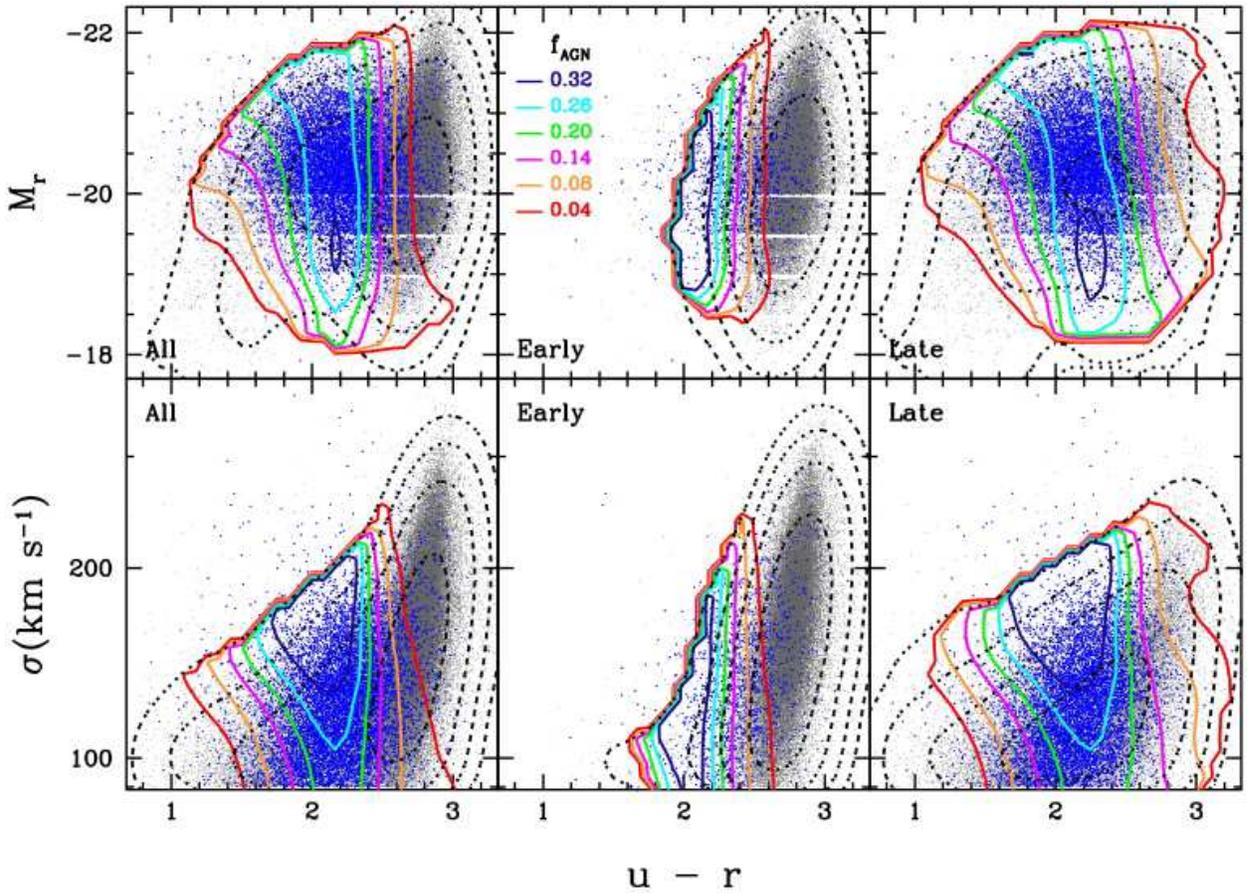}             
\caption{The fraction of AGNs in early-type (solid lines) and late-type galaxies 
(dotted lines) as a function of luminosity ({\it upper}) and velocity 
dispersion ({\it lower}). The sizes are
$\Delta({M_r})$=0.14, $\Delta(log \sigma)=0.02$, and  $\Delta(u-r)=0.08$.
}
\label{fig_urcon}
\end{figure*}
Figure~\ref{fig_lf} presents the number density distributions of 
non-AGN (dashed line) and AGN host (solid line) galaxies as a 
function of luminosity ($M_r$) and velocity dispersion.
Morphology dependence of the distributions is examined by dividing the sample
into early-type (circles) and late-type galaxies (stars).
The shape of luminosity function and velocity dispersion function 
of AGN host galaxies are qualitatively different from those
of non-AGN galaxies. 
From the ratios $\phi(E/L~ AGN)/\phi(all~ AGN)$, one can see that
AGNs reside preferentially in luminous ($M_r \leq -19$) or 
intermediate velocity dispersion ($100<\sigma<200$ km s$^{-1}$) 
late-type galaxies,
although at higher velocity dispersions ($> 200 $ km s$^{-1}$) 
early-type galaxies host comparable number of AGNs compared 
to late-type galaxies.
Note that since the emission line strength roughly scales with black hole
mass (and galaxy mass), detecting AGNs based on the emission line diagnostic is much 
more difficult at low-mass scales, which is reflected in the decreasing 
number density of AGN host galaxies at lower luminosity ($M_r > -19$) or 
lower velocity dispersion ($\sigma$ $<$ 100 km s$^{-1}$).

To understand the coupled dependence of \fAGN\ on $M_r$ and $\sigma$,
we inspect how the \fAGN\ varies in the luminosity versus 
velocity dispersion parameter plane in Figure~\ref{fig_mcon}.
We present galaxies with \sig\ $> 90$ km s$^{-1}$ to reduce systematic 
errors induced by the instrumental resolution of SDSS spectroscopy.

On top of the distributions of the AGN host galaxies (blue dots) 
and all galaxies in the sample (gray dots) superposed are 
the contours of the constant AGN fractions in colored solid lines, 
while the dotted contours represent the number density of all galaxies in the sample. 
The smooth distributions of AGN fraction are obtained from the ratio of the
sum of the weighted number density of AGN host to the sum of the weighted
number density of all galaxies within the smoothing kernel at each bin in the
parameter space. A fixed-size spline-kernel is used to calculate the weights.
Contours are limited to regions with statistical significance above 
$1\sigma$ for the fraction. The parameter bin sizes for making the contours 
are denoted in each figure caption.
A median uncertainty of the \fAGN\ is 5\%, 9\%, and 5\% for
all, early, and late-type galaxies, respectively. 
The \fAGN\ can be interpreted as the probability for a randomly chosen galaxy
to host AGN. As shown in Figure~\ref{fig_mcon}, the \fAGN\ is 
less than $\sim$25\% over all $M_r$ or \sig\ ranges, regardless of morphological type. 
However, we note that \fAGN\ is a lower limit since AGNs with weak emission lines
(S/N $<$ 6) are not included in our AGN sample. 

For the combined sample of early and late-type galaxies in Figure~\ref{fig_mcon} (left panel),
the \fAGN\ increases as host galaxy luminosity increases, 
suggesting that more luminous galaxies are more likely to host AGN. 
In contrast, for given host galaxy luminosity, 
\fAGN\ typically increases with decreasing \sig.
The \fAGN\ dependency on the host galaxy morphological type is clearly seen 
when early-type and late-type galaxies are divided.
In general, \fAGN\ of early-type galaxies is much lower than that of late-type 
galaxies, especially at high luminosity and high \sig.

At fixed luminosity, \fAGN\ of early-type galaxies
is a decreasing function of \sig, indicating that galaxies 
with lower velocity dispersion (\sig $< 200$ \kms) are dominant AGN hosts
among early-types. The decreasing \fAGN\ with increasing \sig\ suggests 
that early-type galaxies with more massive black holes (assuming the 
$M_{\rm BH}-\sigma$ relation)
are less likely to host AGN. Presumably, these most massive galaxies
and their black holes are already well evolved as dormant black holes
in the present-day universe. In contrast, late-type galaxies with intermediate
\sig\ ($\sim$130 \kms) show highest \fAGN.
This is because the fraction of
red galaxies increases as $\sigma$ increases, and
that \fAGN\ is actually a monotonically increasing function of $\sigma$
in the case of late-type galaxies (see \S 4.2).

At fixed \sig, \fAGN\ increases with luminosity for early-type
galaxies, while this dependency becomes weaker for late-type galaxies.
The decrease of \fAGN\ at lower luminosity and lower \sig\ for late-type galaxies
is not straightforward to interpret, however,
we speculate that it could be due to 
the lack of black holes in some fraction of low-mass disk-dominated galaxies or 
the relatively weak AGN activity 
(e.g., low [OIII] luminosity for given Eddington ratios) in these galaxies.

To investigate whether this trend is caused by some selection effect
due to the [OIII] flux limit in the survey as discussed in \S 3.3,
we used a subsample of high-power AGNs with \accrate $> 1$ (\sunit) 
in the calculation of the $f_{\rm AGN}$. 
The fraction of the high-power AGNs is much lower than that of all AGNs. However, 
the overall trend is qualitatively the same and is not due to the [OIII] flux
limit of the sample.

\subsection{Color}
\label{subsec_color}

In this section, we investigate how \fAGN\ is related with the host galaxy 
$u-r$ color, which is an indicator of star formation history or mean stellar age. 
Again, we exclude galaxies with \sig\ $< 90$ km s$^{-1}$ to reduce systematic 
errors induced by the instrumental resolution of SDSS spectroscopy. However, the 
results remain qualitatively the same even when the whole sample is used.

The \fAGN\ dependency on host galaxy color is presented in 
$M_r$ or $\sigma$ versus the $u-r$ color plane (Figure~\ref{fig_urcon}). 
In general AGN host galaxies show a narrower distribution in $u-r$ color than 
non-AGN galaxies, indicating that galaxies with intermediate color 
are dominant AGN hosts. The number density of AGN host galaxies peaks at 
$u-r \sim 2.4$. 
The high fraction of AGN host galaxies in the intermediate-color region between 
the red sequence and the blue cloud is consistent with other studies (e.g. 
Nandra et al. 2007; Schawinski et al. 2007; Martin et al. 2007).
For early-type galaxies, the peak of \fAGN\ is clearly offset from the peak
of number density of non-AGN galaxies. The \fAGN\ of early-type galaxies 
monotonically increases as the galaxies become bluer and reaches 
the maximum at $u-r \sim 2$.
Similarly, for late-type galaxies \fAGN\ is maximum at intermediate color
$u-r = 2 \sim 2.4$.

In contrast to the dependency on color, \fAGN\ of early-type galaxies
does not strongly depend on luminosity or velocity dispersion. 
The middle column of Figure~\ref{fig_urcon} shows that
\fAGN\ of early-type galaxies depends almost entirely on color, implying that
triggering of AGN activity is independent of the luminosity or velocity dispersion
at fixed galaxy color.
The dependence of \fAGN\ on $M_r$ or $\sigma$ found in the previous section is 
merely due to the fact that $M_r$ or $\sigma$ 
is correlated with $u-r$ in the sense that early-type galaxies with 
higher velocity dispersion are redder and that for given \sig, luminosity
is higher for AGN host galaxies than non-AGN galaxies (see Figure 4 upper panel).

Combined with Figure~\ref{fig_mcon}, these results indicate that
\fAGN\ strongly depends on host morphology and $u-r$ color.
For early-type galaxies, it does not much depend on luminosity or velocity
dispersion. 
Thus, one of the necessary conditions for early-type galaxies to trigger
AGN activity is the bluer color, which is an indication of the presence of 
young stellar population and/or cold gas that can be accreted to the black hole.
In contrast, late-type galaxies show weak net dependence of \fAGN\ on 
$M_r$ and $\sigma$ in addition to $u-r$ (right panels in Fig.~\ref{fig_urcon}).
While late-type galaxies with intermediate color
show the highest \fAGN, \fAGN\ at fixed color increases with luminosity
or velocity dispersion. The dependence on luminosity or velocity dispersion
becomes much weaker for the redder late-type galaxies ($u-r > 2.6$), 
as similarly shown among early-type galaxies.
Note that for late-type AGN host galaxies, the range of color distribution becomes
larger with increasing velocity dispersion, implying   
that at lower mass scales AGN activity is more difficult to be triggered and
only occurs in narrower galaxy color ranges.

\subsection{Color gradient}
\label{subsec_colorgrad}

\begin{figure*}
\epsscale{1.1}
\plotone{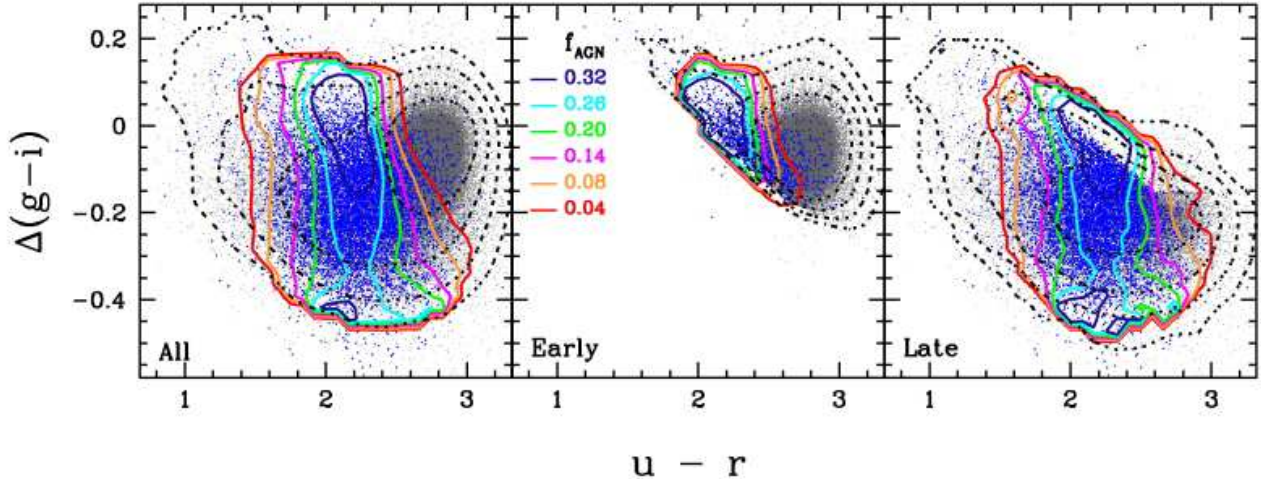}             
\caption{
Distributions of AGN host ({\it blue dots}) and all galaxies
({\it gray dots}) in the $u-r$ color  versus $g-i$ color gradient plane.
The bin sizes are $\Delta(\Delta(g-i))=0.026$ and $\Delta(u-r)=0.08$.
}
\label{fig_urcdcon}
\end{figure*}

We investigate how \fAGN\ depends on the color gradient in 
Figure~\ref{fig_urcdcon}, where the distribution of \fAGN\ is presented 
in the $u-r$ color versus $\Delta(g-i)$ plane.
Note that early-type and late-type galaxies occupy distinct regions 
because galaxy morphology is mainly determined in this plane
(see Figure~\ref{fig_morph}).

For early-type galaxies, the dependence of \fAGN\ on $\Delta(g-i)$ is not
strong.
At $u-r\gtrsim 2.4$, where most of early-type galaxies
are located, the \fAGN\ suddenly drops, indicating that 
typical early-type galaxies are not likely to host AGN.
At bluer colors ($u-r \lesssim 2.4$) where the \fAGN\ is highest,
the \fAGN\ is independent of $\Delta(g-i)$.
The galaxies in this color range have 
on average more positive $\Delta(g-i)$ (i.e. bluer center).
Perhaps, this may indicate the association between AGN and 
blue-core spheroids (e.g. Menanteau et al. 2005) and imply   
that AGN activity is strongly correlated with the amount of cold 
gas or younger stellar population in the bulge.
This trend is also found when the subsample of high-power AGNs ( \accrate $> 1$ \sunit)
is used.

In the case of late-type galaxies, the AGN fraction
has negligible dependence on $\Delta(g-i)$ at fixed color,
probably due to the overall presence of cold gas and star-formation among late-type galaxies. 

\subsection{Light concentration}
\label{subsec_cindex}

\begin{figure*}
\epsscale{1.1}
\plotone{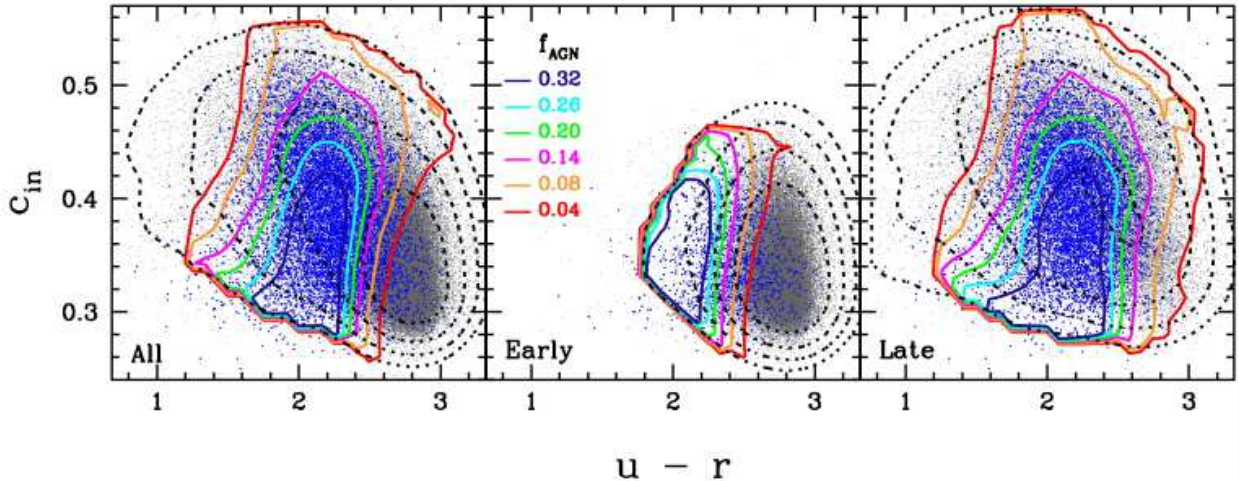}           
\caption{Distributions of AGN host ({\it blue dots}) and all galaxies
({\it gray dots}) in the concentration index versus color plane.
The bin sizes are $\Delta(\Delta(g-i))=0.01$ and $\Delta(u-r)=0.08$.
}
\label{fig_cincon}
\end{figure*}

We investigate how \fAGN\ varies with light concentration ($c_{\rm in}$)
in Figure~\ref{fig_cincon}. 
The dependence of \fAGN\ on $c_{\rm in}$ is relatively weak compared to that
of $u-r$ color. For early-type galaxies, \fAGN\ 
shows only negligible dependence on $c_{\rm in}$.

At $u-r \sim 2$, the \fAGN\ becomes dependent 
on $c_{\rm in}$, especially for late-type galaxies.
Among late-type galaxies with $u-r\lesssim 2.6$,
more concentrated ones have higher \fAGN\ at a given color.
On the other hand, for redder late type galaxies ( $u-r > 2.6$),
\fAGN\ is independent of the light concentrations.
The dependence of \fAGN\ on $c_{\rm in}$ is very similar to that on \sig\ 
as shown in Figure~\ref{fig_urcon} due to the correlation between $c_{\rm in}$ and \sig.
Given the tight correlations between $c_{\rm in}$ and \sig,   
and between \mbh\ and host galaxy bulge mass,
these results may suggest that among late-type galaxies with enough cold gas  
($u-r<2.6$), more bulge-dominated galaxies hosting more massive black hole 
are more likely to trigger AGN activity.
For the redder late-types, which are on average more bulge-dominated,
the amount of cold gas seems more requisite ingredient for AGN activity.
The trend remains the same when we use a subsample with a narrow \sig\ range 
(100 to 170 km s$^{-1}$) in order to separate the effect of $c_{\rm in}$ from 
that of $\sigma$. 

\section{Effects on AGN Power}

In this section, we examine how the host galaxy properties are related
with the power of AGN activity, by adopting \LOIII\
normalized by \mbh\ as an Eddington ratio indicator 
(see Fig.~\ref{fig_OIIIMBH} and \S~\ref{sec_OIIIMBH}). 
First, we compare host morphology with AGN power (\S~\ref{power_morph}).
Then, we investigate how color, color gradient, and concentration index are
related with AGN power (\S~\ref{power_detail}).

\subsection{Host morphology}
\label{power_morph}

\begin{figure}
\epsscale{1.3}
\plotone{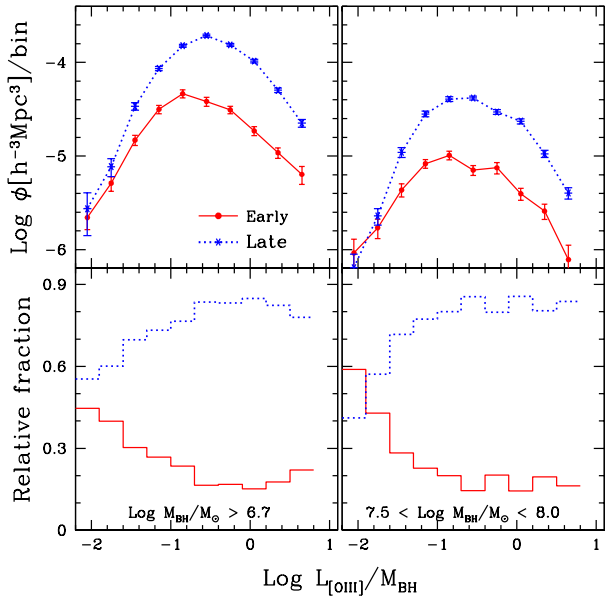} 
\caption{The Eddington ratio distributions ({\it upper}) and
relative fraction of host galaxy morphology ({\it lower})
for AGNs with log $M_{\rm BH}/M_{\odot} > 6.7$({\it left}) and 
AGNs with $7.5<{\rm log} M_{\rm BH}/M_{\odot} < 8$ ({\it right}). 
}
\label{fig_EDD}
\end{figure}

Figure~\ref{fig_EDD} presents AGN number density as a function of AGN power
(\LOIII/\mbh). 
Here, we consider AGNs with log \mbh/$M_{\odot}> 6.7$ (left panels)
since the systematic error due to the instrumental resolution of SDSS spectroscopy
becomes severe at \sig$<$90\kms.
Regardless of host morphology, the Eddington ratio ranges over 3 orders of 
magnitude. The distribution of AGNs hosted by late-type galaxies peaks
at $L_{\rm [OIII]}/M_{\rm BH}$ $\sim$ 0.3 ($L_{\odot}/M_{\odot}$).
For AGNs hosted by early-type galaxies, the peak of the distribution 
is slightly lower by $\sim$0.2 dex. 
The large range of the Eddington ratio distribution of Type II AGNs is consistent 
with that of broad-line AGNs (Woo \& Urry 2002; cf. Kollmeier et al. 2006), 
indicating that AGN power ranges from the Eddington luminosity to 10$^{-2}$ of the 
Eddington limit.      

The number density decreases at the high end of the distribution,
reflecting an intrinsic luminosity function at fixed \mbh.
In contrast, the decrease of the number density at the low end of the
distribution is probably due to the selection effects since weak [OIII] lines 
are hard to detect for a given measurement limit of the [OIII] line flux. 
The number density may keep increasing down to non-AGN level if we include
objects with very low level AGN activity, for which emissions from HII regions 
in the host galaxy dominate over AGN emission lines and/or [OIII] line is 
intrinsically too weak to detect with a few $\AA$ EW detection limit. 
This line of conjecture 
can be supported by the fact that the overall AGN fraction in our sample is less 
than 10\%. This is much lower than the result of the nearby galaxy survey by Ho et al. (1997), 
where many dwarf AGNs with weak emission lines (EW$\sim$0.25\AA) were detected owing to 
the high spatial resolution $\sim$200 pc in their survey. 
Thus, the AGN number density in our
study should be taken as a lower limit since our AGN sample do not include 
very weak emission line AGNs. The intrinsic distribution of the Eddington 
ratio is closely related with the duty-cycle of AGN activity and beyond the scope 
of this paper.

The striking difference between AGNs hosted by galaxies of different morphology 
comes from the overall shape of the Eddington ratio distributions. 
Late-type galaxies are dominant hosts over all AGN power, while the fraction 
of early-type host galaxies increases for low-power AGNs
($L_{\rm [OIII]}/M_{\rm BH}< 0.1$, or $<$ 1\% of the Eddington limit). 

As shown in Figure~\ref{fig_OIIIMBH}, the Eddington ratio range of lower \mbh\ 
objects is biased toward high-power due to the detection limit 
of [OIII] line (i.e. log $L_{\rm [OIII]}/L_{\odot}$ $\sim 5 - 6$). 
Hence, the peak of the distribution observed in the total sample may have 
shifted toward higher Eddington ratios. To avoid this selection effect and 
secure a consistent range of the Eddington ratio, we consider AGNs in a narrow 
\mbh\ range, 7.5 $<$ log $M_{\rm BH}/M_{\odot}< $ 8 
(right panels in Fig.~\ref{fig_EDD}). 
This narrow range of \mbh\ is also useful to avoid any systematic effect 
caused by slightly different \mbh\ distributions of two morphological 
subsamples.
Figure~\ref{fig_EDD} right panels show that the Eddington ratio distributions 
of AGNs in the narrow \mbh\ ranges show the same trend, confirming that  
early-type galaxies host on average less powerful AGNs and that the relative fraction
between early-type and late-type hosts is a strong function of AGN power.

\subsection{Color, color gradient, and concentration}
\label{power_detail}

\begin{figure}
\epsscale{1.2}
\plotone{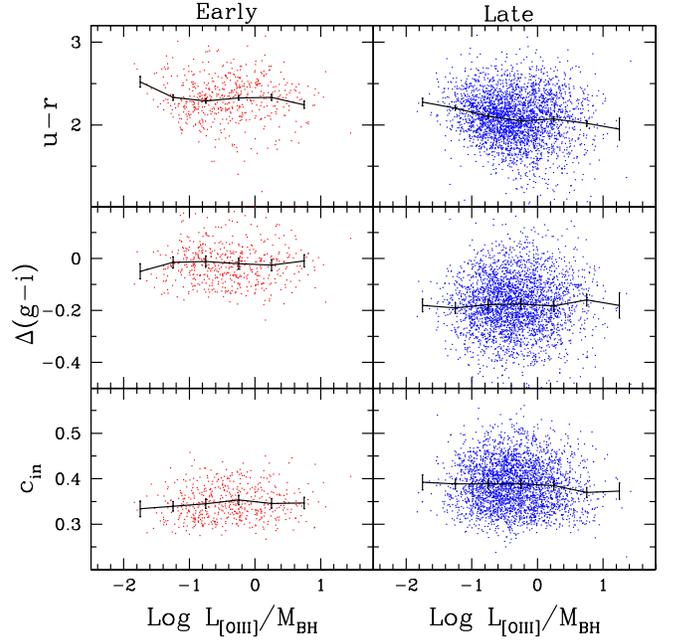}   
\caption{Galaxy properties: $u-r$ color ({\it upper}), color gradient ({\it middle}),
and concentration index ({\it lower}) as a function of the Eddington ratio. 
AGNs with $7 < {\rm log} M_{\rm BH}/M_{\odot} < 8$ are plotted to avoid selection effects.
The error bars include measurement errors of each parameter as well as the scatter in each bin.}
\label{fig_accretion}
\end{figure}

Figure~\ref{fig_accretion} presents host galaxy color, color gradient, and 
concentration as a function of AGN power.
AGNs with $7 < {\rm log} M_{\rm BH}/M_{\odot} < 8$ are 
presented here to minimize selection effects since for lower mass AGNs 
the Eddington ratio distribution is biased 
against low-power objects (see Fig.~\ref{fig_OIIIMBH}).
In this \mbh\ range, the Eddington ratio distribution is not heavily
biased down to $L_{\rm [OIII]}/M_{\rm BH}$ $\sim 0.1$
($\sim$1\% of the Eddington limit),
and the [OIII] luminosity is typically larger than 10$^{6}$ $L_{\odot}$.

Top panels in Figure~\ref{fig_accretion} present the color distribution of
host galaxies as a function of AGN power, showing that low-power AGNs 
are hosted by slightly redder galaxies especially in late-type galaxies,
although at given AGN power there is a spread color distribution.
As AGN power increases from 1\% to 100\% of the Eddington limit,
the average $u-r$ color of late-type galaxies becomes bluer by 
$\sim 0.2$.

In the case of color gradient (middle panels), AGN power do not show strong dependence
since galaxy colors are averaged in the color gradient space. 
A large fraction of blue AGN hosts tend to have bluer inner part 
(as shown in Fig.~\ref{fig_urcdcon}), and have higher Eddington ratios. 
however, this trend seems insignificant in the color gradient space.

The concentration index also shows no strong trend. Among early-type hosts,
slightly less concentrated galaxies host more powerful AGN.
However, as in the case of color gradient, the range 
in the concentration index is large at any given AGN power, indicating that
this dependency is weak. 
As discussed in Section 4, when the colors of galaxies are limited 
to a small range so that $c_{\rm in}$ or $\Delta(g-i)$ becomes a determining factor 
of \fAGN, its dependence on AGN power is slightly improved, however, it 
is still not significant.

\section{Summary and Discussion}

We conservatively identified 11,521 Type II AGNs in a large 
volume- and luminosity-limited sample of 144,940 galaxies with $0.025< z < 0.107$, selected from SDSS DR5,
to investigate how host galaxy properties are related with AGN fraction and power.
The main results are summarized as follows:
\begin{enumerate}
\item
Among other galaxy properties, $u-r$ color shows the dominant dependence of the AGN fraction.
In particular for early-type galaxies, the \fAGN\ does not depend on luminosity
or velocity dispersion at fixed $u-r$ color.

\item AGNs are typically hosted by intermediate-color late-type ($u-r=2\sim2.4$)
and bluish early-type galaxies (peak at $u-r\sim$2.0),
indicating that AGN host galaxies have younger stellar population
than non-AGN galaxies at given luminosity or velocity dispersion.

\item 
AGNs are dominantly hosted by late-type galaxies with intermediate velocity dispersion
($\sim$130 km s$^{-1}$),
while among galaxies with highest velocity dispersion ($\sigma > 200$ km s$^{-1}$),
the number of early-type host galaxies is comparable to that of late-type galaxies.

\item 
Regardless of morphological types, the fraction of AGN increases 
with galaxy luminosity when their velocity dispersions are fixed.
In contrast, at fixed luminosity the \fAGN\ of early-type galaxies 
monotonically decreases as velocity dispersion increases, 
while the \fAGN\ of late-type galaxies peaks at 
intermediate velocity dispersions ($\sigma \sim 130$ \kms).

\item 
The AGN fraction is independent of the color 
gradient in the case of late-type and bluer ($u-r < 2.4$) early-type galaxies. 

\item Among late-type galaxies, more bulge-dominated 
(higher velocity dispersion and more concentrated) galaxies show a higher AGN fraction. 

\item Late-type galaxies are dominant host of AGN of all power (10$^{-3} <$ Eddington ratio)
while the fraction of early-type host galaxies increases at low power ($<1$\% of 
the Eddington limit).
For both morphological types, the Eddington ratio derived from $L_{\rm [OIII]}/M_{\rm BH}$ ranges over 
3 orders of magnitude, indicating various levels of accretion for given \mbh.

\item On average bluer late-type galaxies host more powerful AGNs while
color gradient and concentration index do not show strong dependence on AGN power.

\end{enumerate}

Our results indicate that the requisite ingredient for triggering AGN
activity is massive (high velocity dispersion) bulges and intermediate colors. 
The necessary condition for AGN activity seems mostly different with galaxy morphology.
These results are consistent with a scenario that in the present-day universe,
early-type galaxies with higher velocity dispersion and redder color
are harder to host AGNs since these galaxies already consumed gas at the center or
do not have sufficient gas supply to the central black hole.
In contrast, intermediate-velocity dispersion, intermediate-color, 
and more concentrated
late-type galaxies are more likely to host AGNs. Perhaps, this trend results from
that some fraction of low-mass, blue, and less concentrated late-type galaxies
may not host massive black holes, as predicted by black hole seed models
(e.g. Volontary et al. 2008) or may host very low-power AGNs. The presence of
black holes, either dormant or active, in low-mass disk-dominated galaxies
is currently a topic of active research (see related discussions by Gallo et al. 2008;
Ferrarese et al. 2006)

It is not straightforward to compare our results with previous studies
due to several differences in the sample selection and
analysis. The different schemes we adopted were to match galaxy properties
when we compared AGN host and non-AGN galaxies, and to normalize the AGN luminosity
by black hole mass. While Kauffmann et al. (2003) found that AGNs reside
almost exclusively in massive galaxies, our results show that late-type galaxies
with intermediate-luminosity and intermediate-velocity dispersion are dominant
AGN host in the present-day universe. This different result is probably due to
the fact that our AGN sample is more conservatively defined with a higher S/N cut 
($S/N > 6$) on each emission lines used in flux ratio diagrams, and do not include
very low-power AGNs, which are perhaps more prevalent among massive galaxies.

Although we find strong dependence of AGN activity on host galaxy properties,
we do not find any {\it sufficient} host galaxy property 
to trigger AGN activity since there are more non-AGN galaxies than AGN host galaxies
for any given combination of host galaxy properties. These findings may imply
that large scale environments are crucial for triggering AGN activity
or that local processes at the vicinity of the central black
holes play an important role, especially for the dominant low-luminosity AGNs
in the present-day universe. It is also possible that the life time of AGN is shorter
than that of imprint of triggering mechanisms so that distinct galaxy properties, i.e.,
intermediate color, are still observable although AGN activity finished.
We note that these findings are relevant
only for local galaxies and that host galaxy properties may show 
very different relations with AGN activity at high redshifts.

\acknowledgments

CBP and YYC acknowledge the support of the Korea Science and Engineering
Foundation (KOSEF) through the Astrophysical Research Center for the
Structure and Evolution of the Cosmos (ARCSEC).
JHW acknowledges the support provided by NASA through Hubble Fellowship grant HF-0642621
awarded by the Space Telescope Science Institute, which is operated by the Association of Universities for Research in Astronomy, Inc., for NASA, under contract NAS 5-26555.
We thank Tommaso Treu, Luis Ho, and Matt Malkan for useful discussions. 
We thank the anonymous referee for useful comments.

Foundation, the Participating Institutions, the National Science 
Foundation, the U.S. Department of Energy, the National Aeronautics and 
Space Administration, the Japanese Monbukagakusho, the Max Planck 
Society, and the Higher Education Funding Council for England. 
The SDSS Web Site is http://www.sdss.org/.
The SDSS is managed by the Astrophysical Research Consortium for the 
Participating Institutions. The Participating Institutions are the  
American Museum of Natural History, Astrophysical Institute Potsdam,
University of Basel, Cambridge University, Case Western Reserve University,
University of Chicago, Drexel University, Fermilab, the Institute for
Advanced Study, the Japan Participation Group, Johns Hopkins University,
the Joint Institute for Nuclear Astrophysics, the Kavli Institute for
Particle Astrophysics and Cosmology, the Korean Scientist Group, the 
Chinese Academy of Sciences (LAMOST), Los Alamos National Laboratory,
the Max-Planck-Institute for Astronomy (MPIA), the Max-Planck-Institute
for Astrophysics (MPA), New Mexico State University, Ohio State University,  
University of Pittsburgh, University of Portsmouth, Princeton University,
the United States Naval Observatory, and the University of Washington.

\end{document}